\documentclass{aastex62}

\newcommand\eg{\citep[eg.][]}
\newcommand{\OMM}{\, \textquoteleft Oumuamua}

\graphicspath{{./}{figures/}}

\received{January 1, 2018}
\revised{January 7, 2018}
\accepted{\today}
\submitjournal{ApJL}

\shorttitle{Galactic Panspermia}
\shortauthors{Ginsburg et al.}
\begin{document}

\title{Galactic Panspermia}

\correspondingauthor{Idan Ginsburg}
\email{iginsburg@cfa.harvard.edu}

\author{Idan Ginsburg}
\affil{Institute for Theory and Computation, Harvard University, Cambridge MA 02138, USA}

\author{Manasvi Lingam}
\affiliation{Institute for Theory and Computation, Harvard University, Cambridge MA 02138, USA}

\author{Abraham Loeb}
\affiliation{Institute for Theory and Computation, Harvard University, Cambridge MA 02138, USA}



\begin{abstract}

We present an analytic model to estimate the total number of rocky or icy objects that could be captured by planetary systems within the Milky Way galaxy and result in panspermia should they harbor life. We estimate the capture rate of objects ejected from planetary systems over the entire phase space as well as time. Our final expression for the capture rate depends upon the velocity dispersion as well as the characteristic biological survival time and the size of the captured object. We further take into account the number of stars that an interstellar object traverses, as well as the scale height and length of the Milky Way's disk. The likelihood of Galactic panspermia is strongly dependent upon the survival lifetime of the putative organisms as well as the velocity of the transporter. Velocities between $10-100$ km s$^{-1}$ result in the highest probabilities. However, given large enough survival lifetimes, even hypervelocity objects traveling at over 1000 km s$^{-1}$ have a significant chance of capture, thereby increasing the likelihood of panspermia. Thus, we show that panspermia is not exclusively relegated to solar-system sized scales, and the entire Milky Way could potentially be exchanging biotic components across vast distances.

\end{abstract}



\section{Introduction} \label{sec:intro}

Life is self-serving, with the ultimate aim arguably being survival and consequently propagation \eg{Dawkins95}. 
%Intelligent and technologically advanced civilizations that are long-lived will at some point either need to colonize a new planet, or simply wish to do so. 
The idea that life can self-propagate from one planet to another , {\it panspermia}, has been around for at least thousands of years \citep[for a review see][]{Wesson10,Wick10}. In an address to the British Society for the Advancement of Science in 1871, Lord Kelvin suggested that life could propagate by traveling on meteorites \citep{Thomson-Kelvin}. More than a century later, it was hypothesized that some achondrites were of Martian origin \eg{Bogard83,Smith84}. Subsequent studies confirmed the Martian origin of such meteorites \eg{Nyquist01}, and today over a hundred have been identified\footnote{https://www2.jpl.nasa.gov/snc/}. An impactor of $\sim$ 1 km can accelerate debris to above the Martian escape velocity. Such an impact may result in post-shock temperatures of $\sim$ 1300 K and shock pressure as large as 55 GPa. However, this is not necessarily the case, and studies show that some ejecta are not heated much above 373 K, and thus microorganisms can survive such impacts and be ejected into space \citep{Fritz05}. In fact, throughout the history of the Solar system $\sim 10^9$ rock fragments were ejected from Mars with temperatures not exceeding 373 K \citep{MCW00}.
%Nobel Laureate Svante Arrhenius realized that radiation pressure on spores could propel them from one star to the next \citep{Arrhenius1903}. 

The aforementioned mechanism can in principle spread life across planets. However, there has been much debate about the destructiveness of ultraviolet (UV) radiation, particularly on bacteria. For example, \citet{CrickOrgel} argued that microorganisms are likely to perish in space and that panspermia could only be achieved in artificial spacecrafts via so-called ``directed panspermia". However, some strains of bacteria are extremely resistant to radiation. \citet{Pavlov06} argued that some radioresistant bacteria exhibit such extreme tolerances to UV and ionizing radiation that their origin is likely Mars. Furthermore, studies have shown that spores of the bacteria {\it Bacillus subtilis} survive in space for six years \citep{Horneck94}. Even non-radioresistant bacteria can survive in space so long as they are shielded from UV radiation. Such shielding can be as thin as a few centimeters \citep{Horneck01}. Moreover, bacteria have a survival mechanism where a colony surrounds itself within a biofilm which greatly increases protection against UV and other harmful radiation \eg{Frosler17}.  

With the overwhelming evidence that at least some microorganisms can survive ejection and travel in space, we ask ourselves: {\it is intragalactic panspermia viable?} After a rigorous analysis which is described in Section \ref{SecAM}, we estimate the total number of captured objects by planetary systems in the Milky Way, and determine that there is a parameter space which allows the galaxy to be saturated with life-bearing objects. We discuss this, as well as further implications in searching for signatures of panspermia in Section \ref{SecCon}.

\section{Analytic Model}\label{SecAM}

\subsection{Milky Way}
We first construct a detailed analytical model to estimate the total number of captured objects $N_c$ assuming that they traverse through regions of the Milky Way. By ``captured objects'', we refer to entities with sizes $\gtrsim 0.1$ km, which may include ice-rock objects such as asteroids, comets, planetesimals and planets. The capture rate is dominated by binary star systems \citep{LL18}. We assume that the capture of interstellar objects (per stellar encounter) occurs via the standard theory of gravitational interaction with a stellar binary system \citep{Heg75}; note that ``capture'' in this context refers only to temporary capture since the mean residence time around a given star depends on the configuration of that planetary system.  We can estimate $N_c$ via the following expression:
\begin{equation}\label{Ncexp}
N_c = f_B \int \int \int \int \dot{N} \cdot N_\star \cdot P \cdot f(v)\, d^3 v\,dt,
\end{equation}
where $f_B$ denotes the fraction of extrasolar binaries, and is taken to be around $0.4$ \citep{Lada}. $\dot{N}$ represents the capture rate of interstellar objects per binary system, whereas $N_\star$ is the number of stars spanned by the interstellar objects traveling with a relative velocity $v$ over a time $t$. The term $P$ is the biological survival factor, i.e. the fraction of microbial life that survives during the travel between stellar systems and, in principle, can therefore take part in seeding the new planetary system with life. Lastly, $f(v)$ represents the velocity distribution of the interstellar objects in the local frame of the host star. The integral over $v$ accounts for the velocity distribution of interstellar objects, owing to which $N_c$ serves as the ensemble-averaged total number of capture objects in the Galaxy. The integral over $t$ occurs because the capture rate must be integrated over time to account for the total number of captured objects.

It should be recognized at this stage that $N_c$ yields the number of captured objects over the Milky Way's history as an upper bound on the number of objects that facilitate interplanetary panspermia. The reason being that not all of the temporarily captured objects will actually take part in interplanetary panspermia, since they must be subjected to impacts and the ensuing ejecta must land on a planet with habitable conditions \citep{MCW00,Bur04}. We cannot estimate this probability \emph{a priori} since it depends on the specific properties of the planetary system to which the object has been captured. 

The capture rate $\dot{N}$ can be computed as
\begin{equation}\label{Nstdef}
\dot{N} = n_0 A_c(v) v,
\end{equation}
where $n_0$ is the number density of the interstellar objects and $A_c$ is the capture cross-section. We must now provide suitable expressions for $n_0$ and $A_c$ that are discussed below.

The mathematical details for the capture cross section $A_c$ were first worked out by \citet{Heg75}, and can be expressed as follows  \citep{Val83}:
\begin{equation}\label{siganl}
A_c \propto \left(\frac{a}{1\,\mathrm{AU}}\right)^{-1}\left(\frac{m_1 m_2}{M_\odot^2}\right)^2 \left(\frac{m_1 + m_2}{M_\odot}\right)^{-1}
\left(\frac{v}{1\,\mathrm{km\,s^{-1}}}\right)^{-1}\left(\frac{\sqrt{v^2 + v_b^2}}{1\,\mathrm{km\,s^{-1}}}\right)^{-6}.
\end{equation}
The cross-section $A_c$ depends not only on $v$ but also on the mass of the primary and the secondary ($m_1$ and $m_2$) as well as the orbital velocity of the bound orbit ($v_b$) and the semi-major axis ($a$) of the binary system. If the object is captured into a barely bound orbit, this amounts to taking the limit $v_b \rightarrow 0$ and yields $A_c \propto v^{-7}$. The rationale behind using $v_b$ (instead of dropping it altogether) is explained later.

%We shall assume that the ensemble-averaged value of $A_c$ is comparable to that of the Alpha Centauri A\&B binary system. 
Although there is a wide range of mass distributions and periods for binary systems, the Alpha Centauri system is fairly typical and thus it is apt to assume that the ensemble-averaged value of $A_c$ is comparable to that of Alpha Centauri \citep{Raghavan:2010,Moe17}. Our basic qualitative results remain unaltered when $a$, $m_1$ and $m_2$ are changed. Numerical simulations indicate that the capture cross-section for the Jupiter-Sun binary is about an order of magnitude lower than the analytical estimate \citep{VI82}, and hence $A_c$ for this binary system can therefore be approximated by
\begin{equation}
A_c \sim 10^8\,\left(\mathrm{AU}\right)^2\,\left(\frac{v}{1\,\mathrm{km\,s^{-1}}}\right)^{-1}\left(\frac{\sqrt{v^2 + v_b^2}}{1\,\mathrm{km\,s^{-1}}}\right)^{-6}.
\end{equation}
This capture cross section is about five orders of magnitude higher than the corresponding value for the Jupiter-Sun system \citep{LL18} because of the fact that Jupiter's mass is only $\sim 10^{-3}$ of the stars in the Alpha Centauri A\&B binary system.

The quantity $n_0$ depends on the size $R$ of the object. We will assume that $R \gtrsim 0.1$ km, with the lower bound being the rough size of the recently discovered interstellar object `Oumuamua \citep{Meech17}. On the other hand, for objects with a different minimum size, the cumulative size distribution function must be utilized to adjust $n_0$ accordingly \citep{Dohanyi69}. We will not explicitly introduce the $R$-dependence henceforth to preserve simplicity in notation, except for the final expression (\ref{finale}).

Although this size ($R \gtrsim 0.1$ km) may appear to be on the lower side, we note that the erosion due to dust in the interstellar medium and cosmic rays may affect only the uppermost few meters over a Gyr timespan \citep{HLB17,HLL18}. Thus, given sufficient shielding, it seems plausible that some organic matter (even if not life \emph{per se}) could survive within an interstellar object of size $\gtrsim 0.1$ km over Gyr timescales. Even the transfer of dead organisms or matter, dubbed {\it necropanspermia} and {\it pseudopanspermia} respectively, could be advantageous insofar abiogenesis and nutrient supply is concerned \citep{MCK97,Wesson10,LL17}. 

\begin{figure*}
	\includegraphics[width=1.\columnwidth]{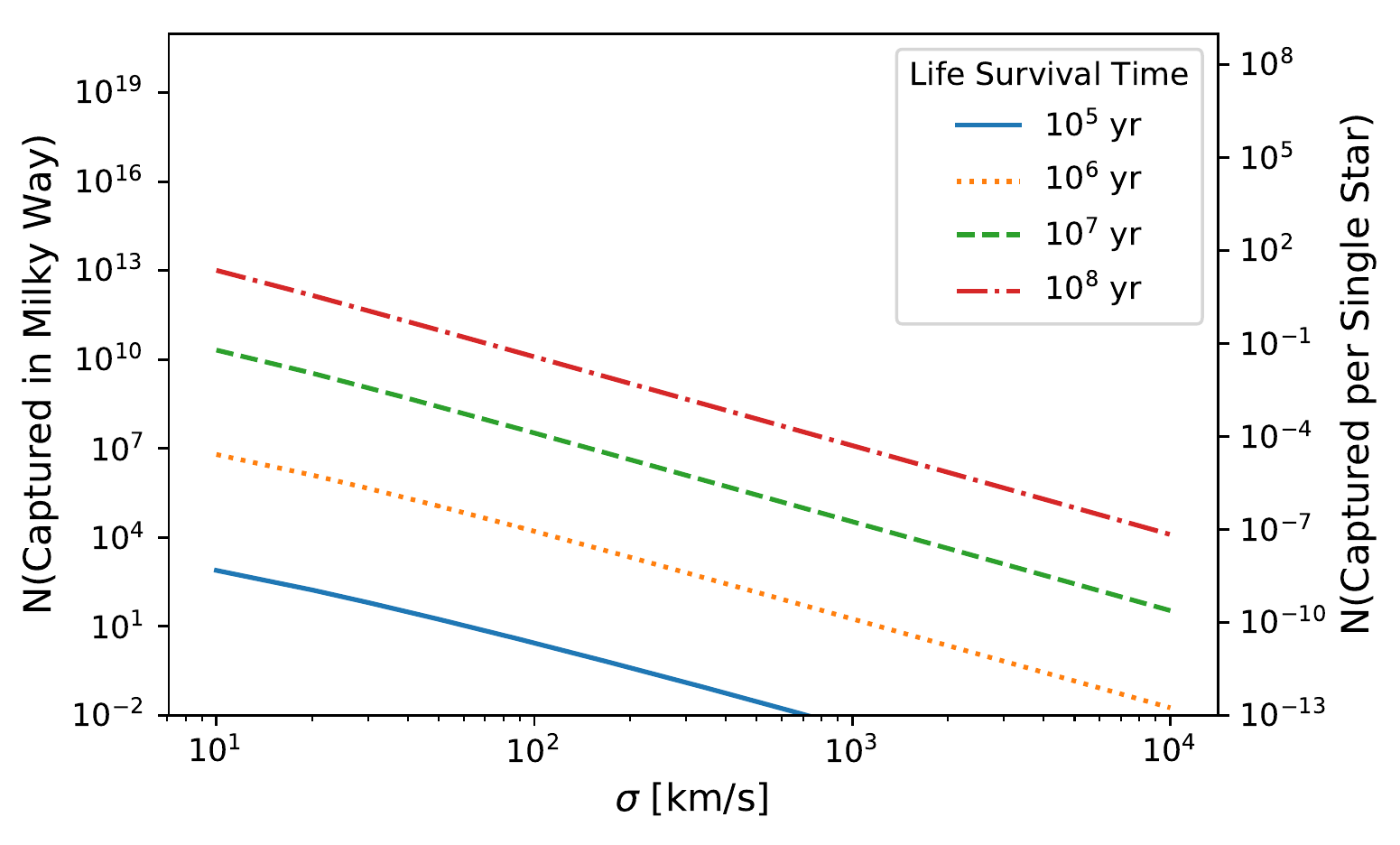}
    \caption{Number of captured \OMM-size objects versus their velocity dispersion. The left column is the number of captured objects within the entire Milky Way. The right column is the number of captured objects per star. We plot the capture rates for biological survival times of $10^5-10^8$ years based on (\ref{IntFin}) with $v_b \sim 20$ km s$^{-1}$. Low survival times naturally result in a lower probability of capture and a lower probability for galactic panspermia. However, bacteria have been shown to survive for at least $10^6$ years. Even at large velocities, potentially life-bearing objects can be captured. Therefore it is possible to seed life across the galaxy and beyond.}
    \label{figA}
\end{figure*}

\citet{DTT18} show the number of \OMM-like objects is $n_0 \sim 10^{15}\,\mathrm{pc}^{-3}$. We have adopted this value as a fiducial estimate. However, the actual number of life-bearing objects is unknown, and could potentially be lower \citep{MCW00} or higher \citep{Belbruno12,Cuk18} by several orders of magnitude. Should future studies enable us to determine better constraints on the number of life-bearing objects, the updated value can be utilized in the subsequent formulae, in particular see (\ref{finale}). Any correction factor should be applied to the vertical axis of Figure \ref{figA} accordingly. Another key assumption we have made is that the total number density of objects available for capture is not dependent on $v$. We have chosen this ansatz since the velocity-dependence of $n_0$ is not empirically constrained. It must be noted that choosing $n_0 \propto v^{-\beta}$ (with $\beta > 0$), or another monotonically decreasing function of $v$, is not expected to affect our results significantly for $\sigma \sim \mathcal{O}(10)$ km s$^{-1}$. However, in the regime $\sigma \gtrsim 100$ km s$^{-1}$, the number of captured objects in Figure \ref{figA} will be lowered accordingly.

By substituting the above numbers in (\ref{Nstdef}), we end up with the capture rate
\begin{equation}\label{Ncapfin}
\dot{N} = 2.3 \times 10^6\,\mathrm{yr}^{-1}\,\left(\frac{\sqrt{v^2 + v_b^2}}{1\,\mathrm{km\,s^{-1}}}\right)^{-6}.
\end{equation}
Next, we will model the velocity distribution of the objects in the local frame of the host binary system by means of a Maxwellian \citep{Has76,VI82},
\begin{equation}
f(v) = \frac{1}{\left(2\pi\sigma^{2}\right)^{3/2}} \exp\left(-\frac{v^2}{\sigma^2}\right),
\end{equation}
where the velocity dispersion is defined as
\begin{equation}\label{sigd}
\sigma^2 = \sigma_\star^2 + \sigma_c^2,
\end{equation}
where $\sigma_\star \sim 10$ km s$^{-1}$ is the characteristic stellar component, and $\sigma_c$ is the typical dispersion for the objects at the point of their initial ejection from the host planetary system. Changing the stellar dispersion to $\sim 30$ km s$^{-1}$ to reflect the value near the solar neighborhood reduces our results by approximately an order of magnitude, but essentially little else is affected. The two dispersions add up since the variance for the sum of two normally distributed variables is the individual sum of their respective variances. Note that $\sigma_c$ constitutes the free parameter of interest in our analysis. 

The next item is the biological survival factor $P$. It is a well-known fact that most cells, and unicellular organisms, exhibit either exponential growth or decay \citep{PTKG}. Thus, we shall assume that $P$, which represents the fraction of microbes that survive within the object over a timescale $t$, could be modeled by
\begin{equation}\label{Ptdef}
P = \exp\left(-\frac{t}{\tau}\right),
\end{equation}
where $\tau$ can be interpreted as a characteristic survival time. However, note that $\tau$ does not represent the survival time for a single microbial organism. Instead, it should be viewed as the typical survival time for the total microbial population hosted within the interstellar object. This survival time will depend on a wide range of properties such as the size and shielding of the interstellar object as well as the radiation and dehydration tolerances of the putative microbes themselves. For instance, based on laboratory experiments, it has been suggested that the fraction of \emph{D. radiodurans} that would survive (with minimal shielding) is $10^{-6}$ over a span of $10^6$ yrs \citep{MCW00,HK10}. Using these values in (\ref{Ptdef}), we obtain a fiducial value of $\tau \sim 10^5$ yrs for polyextremophiles like \emph{D. radiodurans}. Since this estimate assumes only superficial shielding, it is conceivable that the value of $\tau$ could be an order of magnitude higher, especially for the highly radioresistant archaeon \emph{T. gammatolerans}. However, in actuality, we do not know what are the possible values of $\tau$ for extraterrestrial microbes. Thus, we shall leave $\tau$ as another parameter in our analysis. 

The last factor that we wish to estimate is the number of stars $N_\star$ that the interstellar objects span, for a given velocity and duration. This is estimated from
\begin{equation}\label{Nstadef}
N_\star = \int_0^R \int_0^Z n_\star\,\left(2\pi r dr dz\right),
\end{equation}
where the first factor on the RHS is the local stellar density and the second factor is the spatial volume element in the local radial and vertical coordinates of the Galactic disk $(r,z)$. We adopt the axisymmetric ansatz for $n_\star$,
\begin{equation}\label{ndef}
n_\star \sim 1\,\mathrm{pc}^{-3}\,\exp\left(-\frac{z}{H}\right)\exp\left(-\frac{r}{L}\right),
\end{equation}
where $H \sim 0.3$ kpc is the scale height of the thin disk of the Milky Way and $L \sim 3$ kpc is the disk scale length. In (\ref{Nstadef}), $R = v_r t$ and $Z = v_z t$, where $v_r$ and $v_z$ are the radial and vertical components of the velocity. Although we have specified the lower integration limits to be $r=0=z$, this can be easily generalized to other choices. 

After simplification, we find that (\ref{Nstadef}) can be expressed as follows:
\begin{equation}
N_\star \sim 1.7 \times 10^{10}\,\left[1-\exp\left(-\alpha v_z t\right)\right] \left[1-\exp\left(-0.1\alpha v_r t\right)\left(1+0.1 \alpha v_r t\right)\right],
\end{equation}
where $\alpha = 3.3\times 10^{-9}$ and hereafter $v_r$ and $v_z$ are implicitly normalized in units of km s$^{-1}$, whereas $t$ is measured in units of yrs. Lastly, we note that the velocity volume element is given by
\begin{equation}
d^3v = 2\pi v_r dv_r dv_z
\end{equation}

Collecting all of these results and substituting them into (\ref{Ncexp}) leads to
\begin{equation}\label{NcInt}
N_c \sim 3.9 \times 10^{16} \cdot \mathcal{I}_1\left(\sigma,\tau,v_b\right),
\end{equation}
where we have introduced the integral
\begin{equation}\label{Int}
\mathcal{I}_1\left(\sigma,\tau,v_b\right) \equiv \int_{0}^\infty \int_0^{\infty} \int_0^T\, \mathcal{F}\left(v_r,v_z,t\right)\, dv_r\,dv_z\,dt,
\end{equation}
where $T$ represents the total duration over which these objects are assumed to travel. We can choose $T$ to be the age of the Galaxy ($\sim 10^{10}$ yrs), but its actual value does not matter provided that $T \gg \tau$. The integral over $\theta$ has an upper limit of $\pi/2$ because $z > 0$ by definition. Note that the integral is over three variables and not four since the integral over $d\phi$ is straightforward and leads to $2\pi$. The integrand $\mathcal{F}$ is given by
\begin{eqnarray}
\mathcal{F} &\equiv& \frac{1}{\sigma^3}\frac{v_r}{\left(v_r^2 + v_z^2 + v_b^2\right)^3}\exp\left(-\frac{v_r^2 + v_z^2}{\sigma^2}\right)\exp\left(-\frac{t}{\tau}\right) \\
&& \quad \times  \left[1-\exp\left(-\alpha v_z t\right)\right] \left[1-\exp\left(-0.1\alpha v_r t\right)\left(1+0.1 \alpha v_r t\right)\right] \nonumber
\end{eqnarray}
After undertaking the integration with respect to $T$ and taking the limit of $T/\tau \gg 1$, we end up with
\begin{equation}\label{Nccapfin}
N_c \sim 1.4 \times 10^{-11} \cdot \mathcal{I}_2\left(\sigma,\tau,v_b\right),
\end{equation}
where the integral $\mathcal{I}_2$ equals
\begin{equation}\label{IntFin}
\mathcal{I}_2\left(\sigma,\tau,v_b\right) \equiv \frac{\tau^4}{\sigma^3} \int_0^\infty \int_0^\infty \frac{v_r^3 v_z \mathcal{G}\left(v_r,v_z\right)}{\left(v_r^2+ v_z^2 + v_b^2\right)^3}\exp\left(-\frac{v_r^2 + v_z^2}{\sigma^2}\right)\,dv,
\end{equation}
and the function $\mathcal{G}\left(v_r,v_z\right)$ is defined to be
\begin{equation}
\mathcal{G} \equiv \frac{3 + \tau^2 \alpha^2 \left(v_z + 0.1 v_r\right)^2 + \tau \alpha \left(3 v_z + 0.4 v_r\right)}{\left(1+\alpha \tau v_z\right)\left(1+0.1\alpha \tau v_r\right)^2\left(1+\alpha \tau v_z + 0.1\alpha \tau v_r\right)^2}    
\end{equation}
It should be noted that (\ref{IntFin}) formally diverges when $v_b = 0$. Hence, it becomes necessary to work with a \emph{finite} value of $v_b$ (that can be arbitrarily small) for the sake of calculating the integral. Secondly, (\ref{IntFin}) displays a an approximately linear dependence on $v_b$ in the regions of interest.

As the integral (\ref{IntFin}) cannot be evaluated analytically, it is more instructive to study the limiting regimes using the theory of asymptotic integrals \citep{BO78}. First, let us consider the case where $\tau$ is very small, leading to
\begin{equation}\label{I2tzero}
\lim_{\tau \rightarrow 0}\mathcal{I}_2 \propto \tau^4.
\end{equation}
As $\tau$ approaches zero, we would intuitively expect $N_c$ to vanish since all microbes die out, and this can be verified from (\ref{I2tzero}). Similarly, taking the formal limit of large $\tau$ yields
\begin{equation}
\lim_{\tau \rightarrow \infty}\mathcal{I}_2 \propto \tau.
\end{equation}
Thus, based on the above considerations, $N_c$ may possess an approximate power-law dependence on $\tau$ with the corresponding exponent lying between $1$ and $4$. As $N_c$ is strongly sensitive to the value of $\tau$ in the more realistic regime of relatively small $\tau$, it is clear that the survival lifetime of a population will play a major role in determining the number of captured life-bearing objects. It can be readily verified that (\ref{IntFin}) is a monotonically increasing function of $\tau$ indicating that a longer survival lifetime leads to a higher number of captured objects. 

Next, let us consider the different regimes of $\sigma$. The limit $\sigma \rightarrow 0$ is not mathematically allowed, because $\sigma$ has a lower bound of $\sigma_\star \sim 10$ km s$^{-1}$ as seen from (\ref{sigd}). Instead, upon evaluating how (\ref{IntFin}) behaves when $\sigma$ is large, we find
\begin{equation}\label{eq20}
\lim_{\sigma \rightarrow \infty}\mathcal{I}_2 \propto \sigma^{-3}.
\end{equation}
In the regime of interest, i.e. $\sigma \gg 1$ km s$^{-1}$, we find that the number of captured objects becomes a power-law with an exponent of $-3$. Clearly, the number of captured objects becomes vanishingly small for high values of $\sigma$, which is expected. From (\ref{IntFin}), it can be verified that $N_c$ decreases monotonically with $\sigma$. This trend is consistent with the qualitative picture in which higher speeds (velocity dispersions) lead to fewer objects being captured.

The total number of interstellar objects which can be captured by a Milky Way-like galaxy are given in (\ref{IntFin}) which can be further simplified when $\tau \lesssim 10^6$ yrs and $\sigma \gtrsim 100$ km s$^{-1}$. In this regime, the total number of captured objects can be estimated by,
\begin{equation}\label{finale}
N_c \sim C\left( \frac{\tau}{10^6\,\mathrm{yr}}\right)^4 \left(\frac{\sigma}{100\,\mathrm{km\,s^{-1}}}\right)^{-3}\left(\frac{R}{0.1\,\mathrm{km}}\right)^{-2.5},
\end{equation}
where $R$ is the radius of the captured object, and $C$ is a constant that comes from numerically integrating (\ref{IntFin}). For a Milky Way-like galaxy, $C \sim 10^{5}$. The second term arises from (\ref{I2tzero}), and the third term from (\ref{eq20}). The last term has been introduced to account for the fact that the number density of interstellar objects available for capture ($n_0$) depends on $R$, and hence we have introduced the distribution for $n_0$ based on the theoretical prescription derived in \citet{Dohanyi69}. Although we use (\ref{finale}) to calculate the number of captured objects within the Milky Way, it can in principle be used for numerous other systems. In regions of parameter space other than those elucidated earlier, (\ref{finale}) overestimates the actual value by a few orders of magnitude and therefore the correct values can be read off from Figure \ref{figA}.

An interesting consequence of Figure \ref{figA} is that the number of Europa-sized objects ($R \sim 1000$ km) will be
\begin{equation}\label{Europa}
N_c \sim 10^3\left(\frac{\sigma}{10\,\mathrm{km\,s^{-1}}}\right)^{-3}.
\end{equation}
Here, we use $\tau \sim 10^8$ yrs and the fact that the cumulative size distribution scales as $R^{-2.5}$, because radiogenic heating is expected to sustain a subsurface ocean for Europa-sized objects for at least this duration of time \citep{LL17c}. Hence, it is conceivable that the characteristic survival time of putative subsurface microbial biospheres would also be commensurate with this value of $\tau$. From the above equation, it is apparent that only $\sim 10^3$ life-bearing Europa-sized objects could have been captured via panspermia in the Galaxy, provided that $\sigma \sim 10$ km s$^{-1}$. However, it is possible that $\tau \sim 10^9$ yrs \citep{spohn03}, in which case the number of life-bearing Europa-sized objects increases to $\sim 10^5$. 

In comparison, if we consider approximately Enceladus-sized life-bearing objects ($R \sim 100$ km), for $\tau \sim 10^8$ yrs and $\sigma \sim 10$ km s$^{-1}$, we obtain $N_c \sim 10^6$. Changing the timescale to $\tau \sim 10^9$ yrs yields a total of $N_c \sim 10^8$ Enceladus-sized life-bearing objects. Finally, selecting $\tau \sim 10^9$ yrs and $\sigma \sim 10$ km s$^{-1}$ for Earth-sized objects ($R \sim 6000$ km), we find $N_c \sim 1000$. Thus, our analysis suggests that $\sim 1000$ life-bearing exo-Earths have been captured in our Galaxy, but this number must be interpreted with due caution since the cumulative size distribution used herein may not be entirely accurate for planet-sized objects.

%\begin{figure}
%\includegraphics[width=\columnwidth]{Panspermia6.pdf}
%   \caption{Similar to Figure \ref{vfig}, this figure represents the captured objects in a typical globular cluster.}
%   \label{tfig}
%end{figure}

\section{Conclusions}\label{SecCon}
%\OMM

There is no doubt that rocky material can easily be exchanged between nearby planets, such as Mars and Earth or the planets of the TRAPPIST-1 system \citep{LiLo}. Thus, nearly all papers on panspermia focus on interactions within a Solar system (see, however, \citet{Adams05,Li15,Ling16} and references therein). In this paper we conclusively show that panspermia is viable on galactic scales. Figure \ref{figA} shows that objects with lower velocities are in general far more likely to be captured.  
However, for sufficiently long biological survival lifetimes, the probability of capturing a life-bearing object of $v > 100$ km s$^{-1}$ can be significant. 
In particular, interactions in the the Galactic Center (GC) between the supermassive black hole and a stellar binary can accelerate stars to thousands of km s$^{-1}$ \citep{Hills:1991, Ginsburg:1,Ginsburg:2,GuillochonL15,LoebG15,FragioneI}. \citet{Ginsburg:4} showed that the same mechanism can accelerate planets up to $\sim 10^4$ km s$^{-1}$. Assuming planetary systems have asteroids and comets, dynamical interactions with the black holes can eject them at extreme velocities, and thus they may traverse the entire radius of the Milky Way in $\sim 10^6$ yr. Such hypervelocity objects can become intergalactic. However, the capture probability for an intergalactic object is extremely low. If bacteria and other possible extremophiles have sufficiently long survival lifetimes, the GC can act as an engine for panspermia and seed the entire galaxy. 

While the exact survival lifetimes of bacteria is unknown, it is clear that bacteria can survive for at least millions of years \citep{Bidle07}. More complex organisms have been shown to be highly resilient as well. Some species of the phylum \emph{Tardigrada} survived for days in the vacuum of space even when unshielded from radiation \citep{Erdmann17}. Recently, two nematodes (roundworms) were revived after being in cryobiosis for an estimated (3-4)$\times10^4$ years \citep{Worms18}. Thus, although panspermia appears most likely to transfer bacteria and other microbes, it may also transfer more complex organisms. Consequently, viruses will invariably be a part of panspermia for the following reasons.

Biological studies strongly suggests that viruses have long been an influential driving mechanism for the evolution of life on Earth, including human evolution \eg{Koonin06,Durz15,Enard16}. Every place where there is life on Earth, there are also viruses. An integral part of the biosphere, viruses are by far the most prolific entities on Earth with the oceans harboring $\sim 10^{30}$ viruses \citep{Suttle05}. Whether a virus falls under the definition of life has long been debated \citep{Forterre10}. Viruses are capable of Darwinian evolution, but also require a living cell for replication. Regardless, viruses are prolific and can survive in extreme environments such as Arctic conditions, around hydrothermal vents where temperatures can exceed 673 K, and even alkaline lakes \citep{LeRomancer07}. It is extremely likely that bacteria ejected from the Earth via impacts carried with them bacteriophages. Based on the work in this paper, we conclude that viruses could certainly propagate across the Milky Way. While the direct detection of extraterrestrial viruses on an exoplanet is not currently feasible, indirect detection through biosignatures might be possible via geochemical signatures \citep[see][]{Berliner18}.

There are a number of possible ways to detect extrasolar objects that could potentially contain biotic material. Arguably the simplest method is simply detecting an object passing through the solar system. 
The first such fortuitous encounter was with 1I/2017 U1 (\OMM), which has a radius of $\sim 0.1$ km and external velocity $v \sim 25$ km s$^{-1}$ \citep{Meech17}. Unfortunately, at any given time there is likely only $\sim 1$ such objects within 1 au of the Sun \citep{Seligman18}. 
However, \citet{LL18} demonstrated that there may exist $\sim 10^3$ \OMM-like captured objects currently within our Solar system which can be detected in principle by analyzing oxygen isotope ratios and the abundances of 2- and 3-carbon molecules. Furthermore, captured objects may have unusual orbits with large inclinations and high eccentricities \citep{Namouni18}. If it can be shown that such objects are dissimilar in composition and yet extrasolar, this would be evidence, albeit not wholly conclusive, for panspermia on galactic scales. 

\citet{Lin15} used a statistical model to argue that detecting a group of biologically active planets interspersed with areas where life is rare or nonexistent can be nearly irrefutable evidence for panspermia. 
If panspermia operates on galactic scales, it may be that there are little or no habitable exoplanets without life. Consequently, if a survey of exoplanetary atmospheres across the Milky Way found that nearly all habitable planets had biosignatures, this would be strong evidence for galactic panspermia. Methods for identifying panspermia across different planetary systems include the possibility of searching for signatures of homochirality via circularly polarized radiation, and determining the putative lifeforms have the same chirality \citep{LL17}.

Intelligent and technologically sophisticated species that are long-lived will at some point either need to settle a new planet, or simply wish to do so (directed panspermia). Distinguishing between natural panspermia and directed panspermia will be challenging. However, detecting biosignatures on a planet that is not within a habitable zone might be indirect evidence for directed panspermia. Such an argument is significantly strengthened if the same signature is detected on a nearby habitable planet. It is also conceivable that some \OMM-like objects are artificial in nature. Radio observations of \OMM \,\,showed no unusual radio transmissions \citep{ES18,TK18,HarpOM}. Considering the vast number of objects which are expected to be captured (see Figure \ref{figA}) finding signs of extraterrestrial technology will be difficult, although a worthwhile search.

\section*{Acknowledgements}

We thank John Forbes, Sownak Bose, Max Moe, Rahul Kannan and John ZuHone for useful discussions. This work was supported in part by the Breakthrough Prize Foundation, Harvard University, and the Institute for Theory and Computation. 

%%%%%%%%%%%%%%%%%%%%%%%%%%%%%%%%%%%%%%%%%%%%%%%%%%



%\bibliographystyle{aasjournal}
%\bibliography{PAN.bib}

\end{document}